\documentclass[aps,rmp,groupedaddress]{revtex4}

\usepackage{color}

\newif\ifpdf
\ifx\pdfoutput\undefined
\pdffalse 
\else
\pdfoutput=1 
\pdftrue
\fi

\ifpdf
\usepackage[pdftex]{graphicx}
\else
\usepackage{graphicx}
\fi
\usepackage{hyperref}

\begin{document}

\ifpdf
\DeclareGraphicsExtensions{.pdf, .jpg}
\else
\DeclareGraphicsExtensions{.eps, .jpg}
\fi

\def\hslash{\hbar}
\def\imag{i}
\def\grad{\vec{\nabla}}
\def\div{\vec{\nabla}\cdot}
\def\curl{\vec{\nabla}\times}
\def\DDt{\frac{d}{dt}}
\def\ddt{\frac{\partial}{\partial t}}
\def\ddx{\frac{\partial}{\partial x}}
\def\ddy{\frac{\partial}{\partial y}}
\def\lap{\nabla^{2}}
\def\divv{\vec{\nabla}\cdot\vec{v}}
\def\gradS{\vec{\nabla}S}
\def\vvec{\vec{v}}
\def\wc{\omega_{c}}
\def\<{\langle}
\def\>{\rangle}
\def\Tr{{\rm Tr}}
\def\Csch{{\rm csch}}
\def\Coth{{\rm coth}}
\def\Tanh{{\rm tanh}}
\def\g2{g^{(2)}}
\definecolor{macblue}{rgb}{0.19,0.38,1}
\definecolor{orange}{rgb}{1,0.4,0}
\definecolor{bgblue}{rgb}{0.1,0.1,1}
\definecolor{red}{rgb}{1,0,0}
\definecolor{green}{rgb}{0,1,0}
\definecolor{blue}{rgb}{0,0,1}


\preprint{Prepared for CiSE, Computing in Science and Engineering IEEE/AIP}

\title{Quantum Mechanics with Trajectories:
Quantum Trajectories and Adaptive Grids}


\author{Robert E. Wyatt}
\email[email: ]{cman041@aurora.tacc.utexas.edu}
\affiliation{Department of Chemistry and Biochemistry,
University of Texas,
Austin, Texas 78712}

\author{Eric R. Bittner}
\affiliation{Department of Chemistry and Center for Materials Chemistry\\
University of Houston,
Houston, Texas 77204}
\email[email:]{bittner@uh.edu}
\homepage[web page: ]{http://k2.chem.uh.edu/bittner}

\date{\today; Prepared for CiSE, Computing in Science and Engineering IEEE/AIP}

\begin{abstract}
Although the foundations of the hydrodynamical formulation of quantum mechanics were laid over 50 years ago, it has only been within the past few years that viable computational implementations have been developed.  One approach to solving the hydrodynamic equations uses quantum trajectories as the computational tool. The trajectory equations of motion are described and methods for implementation are discussed, including fitting of the fields to gaussian clusters.  
\end{abstract}


\maketitle

\section{Introduction}

The concept of a path or trajectory plays a central role in our understanding of the motion of objects and fluids.  Much like a route traced on a road map, a trajectory tells us where an object started, where it goes, and how it gets there.  There may be alternate routes, some more likely than others.  Hence, analysis of a set of trajectories provides us with an intuitive tool for understanding possible complex dynamics.  Macroscopic objects obey NewtonÕs equation of motion,  $m\ddot{q}=f(q(t))$, 
where $q$  is the position of the particle at time $t $ and $f(q(t))$ is the force acting on it.  
Given values for both position and velocity at time $t$, we can compute the trajectory that the object will follow and as a result, we can predict with certainty where the object will wind up.  

However, at the atomic and molecular level where objects obey the rules of quantum mechanics, NewtonÕs equations of motion are no longer strictly valid and the concept of a unique trajectory given a set of initial conditions becomes murky at best.  This is because, fundamentally, quantum mechanics is non-local, an issue to which we will return later.  In addition, the Heisenberg uncertainty principle states that when measurements are made, we cannot simultaneously determine with infinite precision the exact position {\em and} velocity of a quantum particle, 
although in principle this is possible in Newtonian mechanics.  
Consequently, it seems as though we cannot speak in terms of the unique trajectory followed by a quantum mechanical object. 

However, even in quantum mechanics, Feynman~\cite{ref1} showed that we can talk in terms of paths, in fact, ensembles of them.  For example, if we specify two fixed end-points, $q(0)$ and $q(t)$,  then we can compute the probability of a particle starting at $q(0)$ and winding up at $q(t)$ by summing over all possible paths connecting the two points (these paths include the classical trajectory linking these points, if there is one) and weighting each path by the complex-valued factor $\exp(iS/\hbar)$ , where $S(t)$ is the classical action integral 
 		\begin{eqnarray}
S(t) = \int_0^t \frac{1}{m}\dot{q}^2 - V(q(t)) dt
\end{eqnarray}
in which the integrand is the classical Lagrangian. What we lose here is any indication of exactly which path the particle actually follows, although we can say that some paths are more likely than others, especially those close to the paths predicted by classical mechanics.  In order to make a prediction, it is as if the particle must explore all possible routes between the two end points.  

Another way that we can describe the motion of quantum mechanical objects is via the hydrodynamic formulation of quantum mechanics, which was first introduced in the late 1920Õs and later explored and extensively developed by Bohm~\cite{ref2}
 starting in the early 1950Õs.  Here we will define an ensemble of quantum trajectories, each with a {\em precisely defined}
 coordinate and velocity that uniquely characterizes the dynamical evolution of a quantum system.  However, we still are not free to talk about independent trajectories as in classical mechanics; quantum trajectories are coupled together and evolve as a {\em correlated ensemble}. 
This correlation is a unique feature of quantum dynamics and is expressed as a non-local potential in the hydrodynamic formulation.

In this Article, we will explore the use of the hydrodynamic viewpoint of quantum mechanics to design new computational tools for predicting the evolution of quantum systems. Within the past few years, starting in 1999, it has become possible to directly solve the quantum hydrodynamic equations to {\em predict}
 the space-time dynamics of elements of the probability fluid.\cite{ref3,ref4,ref5} Elements of this quantum fluid are linked through the Bohm quantum potential, denoted $Q$, which is computed on-the-fly as the equations of motion are integrated to generate the hydrodynamic fields. The quantum potential introduces all quantum features into the dynamics, including interference effects, barrier tunneling, zero point energies, etc. The only approximation made in solving the hydrodynamic equations involves the use of a relatively small number of fluid elements.  The equations of motion for these fluid elements are expressed in the moving with the fluid, Lagrangian, picture of fluid flow.

One implementation of these ideas, referred to as the {\em quantum trajectory method} \cite{ref3}  (QTM),
 has been used to 
predict and analyze the dynamics of wavepackets in a number of scattering problems. An approach similar to the QTM is the quantum fluid dynamic method (QFD)\cite{ref5}. Within the past two years, there has been a surge of 
interest in the development and application of trajectory methods for solving both the time-dependent
 Schr{\"o}dinger equation and density matrix equations of motion. Novel quantum trajectory methods for evolution of the reduced density matrix in both non-dissipative and dissipative systems have also been developed\cite{ref6}. In order to circumvent computational problems associated with the propagation of Bohmian trajectories (especially in regions where nodes develop) adaptive grid strategies have been recently explored.\cite{ref4,ref7,ref8}

	In Sec. II of this article, the Bohmian formulation of the hydrodynamic equations of motion will be reviewed, and computational implementations will be described.  Recent studies that employ cluster modeling to calculate the quantum potential will also be presented in this section~\cite{ref9}.  Section III continues with a discussion of adaptive dynamic grid techniques, and the transforms of the hydrodynamical equations appropriate for moving grids are presented.  Finally, some suggestions for further study are presented in Sec.IV.
		  
\section{Implementation of the quantum hydrodynamic equations}

\subsection{The equations of motion}

       In this Section, the equations needed to implement the quantum hydrodynamic formulation will be reviewed.   This formulation is initiated by substituting the amplitude-phase decomposition of the time-dependent wavefunction, 
$\psi(y,t) = R(y,t)\exp[i S(y,t)/\hbar)$, into the time-dependent  Schr{\"o}dinger equation,
\begin{eqnarray}
i\hbar\frac{d\psi}{dt} = \left(-\frac{\hbar^2}{2m}\nabla^2 + V\right)\psi
\end{eqnarray}
In terms of $R$ and $S$, the probability density and the local flow velocity are given by  $\rho= |\psi|^2 = R^2$ and 
$v= j/\rho$ , where $j$ is the probability current. The Lagrangian form of the hydrodynamic equations of motion resulting from this analysis are given by:
\begin{eqnarray}
\frac{d\rho}{dt} = -\rho\nabla\cdot v
\end{eqnarray}
\begin{eqnarray}
\frac{dv}{dt}= -\frac{1}{m}\nabla(V+Q)
\end{eqnarray}
in which the derivative on the left side is appropriate for calculating the rate of change in a function along a fluid trajectory. Equation (3) is recognized as the continuity equation and Equation (4) is a Newtonian-type equation in which the flow acceleration is produced by the sum of the classical force, $f_c = -\nabla V$,
 and the quantum force is $f_q=-\nabla Q$, where $V$ is the 
potential energy function and $Q$ is quantum potential \cite{ref2}.  
The quantum potential measures the curvature induced internal stress and is given by
\begin{eqnarray}
Q(y,t) = -\frac{\hbar^2}{2m}\frac{1}{R(r,y)}\nabla^2R(y,t) = -\frac{\hbar^2}{2m}\rho^{-1/2}\nabla^2\rho^{1/2}
\end{eqnarray}
Computation of the quantum potential is frequently rendered more accurate if derivatives are evaluated using the amplitude
$C = \log(R)$
 ($C$ is referred to as the $C$-amplitude).  In terms of derivatives of this amplitude, the quantum potential is
$Q = -\hbar^2 (\nabla^2 C + (\nabla C)^2)/2m$.

       Closure of the set of dynamical equations is obtained by introducing the equation for the action function,
\begin{eqnarray}
\frac{dS}{dt} = \frac{1}{2m}(\nabla S)^2 - (V+Q) = L_q
\end{eqnarray}
This equation, referred to as the {\em quantum Hamilton-Jacobi equation}, relates the change in the action to the quantum Lagrangian.

	       The quantum trajectories obey two important noncrossing rules: (1) they cannot cross nodal surfaces (along which $\psi=0$ ); (2) they cannot cross each other.  (In practice, because of numerical inaccuracies, these conditions may be violated.)  The quantum trajectories are thus very different from the paths used to assemble the Feynman propagator. 

      {\em No approximations were made in deriving these equations from the time-dependent  Schr{\"o}dinger equation. }
 However, in order to generate solutions to these equations, an approximation will be made.  The initial wavepacket will be subdivided into $N$ fluid elements and the equations of motion will be used to find the hydrodynamic fields along the trajectories followed by these elements.  As time proceeds, the fluid elements evolve into an unstructured mesh, and this presents problems for accurate derivative evaluation. For this purpose, least squares methods may be used.\cite{ref3,ref4} 

\begin{figure}[t]
\includegraphics{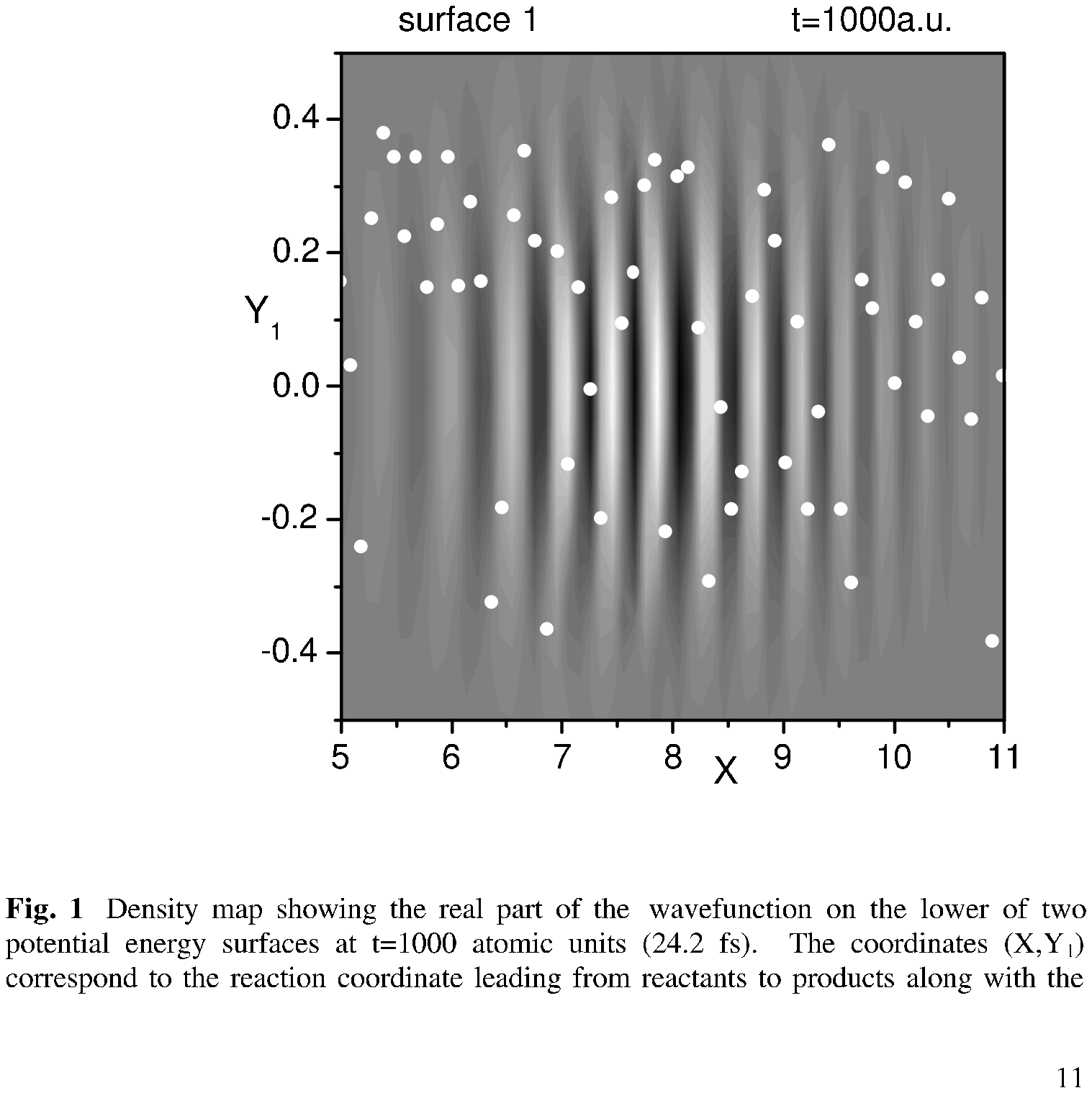}
\caption{Density map showing the real part of the wavefunction on the lower of two potential energy surfaces at $t=1000$ atomic units (24.2 fs).  The coordinates $(X,Y_1)$
 correspond to the reaction coordinate leading from reactants to products along with the first of 10 vibrational coordinates.
  On each potential surface, 110 quantum trajectories were propagated, and some of these are shown by the large white dots.  
The information carried along these scattered trajectories was used to build the wavefunction and this function was later interpolated onto a uniform mesh for plotting purposes.}\label{Fig1}
\end{figure}

	From the values of $C$ and $S$ carried by each evolving fluid elements, 
the complex-valued wavefunction may be synthesized.  An example is shown in Fig.~\ref{Fig1}: this wavepacket has just propagated downhill on an 11 degree-of-freedom potential surface This complicated, oscillating wavefunction (only the real part of the wavefunction is shown) was synthesized from the information carried along 110 quantum trajectories.  It is remarkable that this complicated structure can be built from the information that is propagated along so few quantum trajectories.

\subsection{Multidimensional dynamics with cluster modeling}

For some systems, including those of low dimensionality, the methods introduced in the previous section allow us to solve the quantum hydrodynamic equations of motion to high accuracy.  For higher dimensional systems, such as atomic clusters or larger molecules, an extremely accurate description of the quantum motion is very difficult to obtain and perhaps more difficult to understand.  In this section, we will discuss how the hydrodynamic/Bohmian method can be used to develop an approximate statistical approach for high dimensional systems.\cite{ref9}

We can specify the configuration of a collection of particles by a set of coordinates, 
$R(t) = \{r_1(t),r_2(t)\cdots,r_N(t)\}$. 
The probability of finding one of these configurations is given by the product of the volume element $dV$ weighted by the probability density $\rho(R(t))$. 
In essence, each configuration is an element in an ensemble of possible configurations and the act of measurement pulls one of these configurations out of the hat.  In quantum mechanics, the probability density is related to square amplitude of the quantum wave function, $\psi(R,t)$, Finally, we apply the Bohmian postulate that the time evolution of a configuration, $r_n$, 
is given by \cite{ref2}
\begin{eqnarray}
\dot{r}_n =\left. \frac{1}{m}\nabla_n S(R,t)\right|_{R = r_n}
\end{eqnarray}
where $S$ is the action function for the quantum wave function as described in the previous section. In this section, we discuss how one can use a Bayesian statistical approach to {\em approximate} $Q$ by assuming that the
 density can be cast as a superposition of gaussian product states.

Any statistical distribution can be written as a superposition of a set of $M$ clusters [10], 
\begin{eqnarray}
\rho(R) = \sum_{m=1}^M p(R,c_m)
\end{eqnarray}
where $p(R,c_m)$ is the probability of finding the system in configuration $R$ and being in the $m$-th cluster.  
We then use Bayes\cite{ref10} theorem to break this joint probability into a conditional probability $p(R|c_m)$ 
which tells us the probability that a configuration is a member of $R$ {\em knowing} that it is also a member of the $m$-th 
cluster and a marginal probability, $p(c_m)$ which gives the likelihood of being in the $m-th$ cluster. 
We then pick a functional form for the conditional probabilities by writing 
\begin{eqnarray}
p(R|c_m) = \frac{||{\bf c}_m^{-1}||^{1/2}}{(2\pi)^{3N/2}} \exp\left[-(R-\mu_m)^T{\bf c}_m^{-1}(R-\mu_m)\right],
\end{eqnarray}
where $c_m$ is the $N \times N$ dimensional covariance matrix and $\mu_m$ 
is the center of the gaussian in $N$ spatial dimensions.  
To determine the coefficients of the gaussians, we maximize the log-likelihood, ${\cal L}$, of a given trial set of gaussians actually corresponding to the data by taking the variation  $\delta {\cal L} = 0$. 
 Furthermore, since the density is now represented as a superposition of gaussians, it is a straightforward task to compute the quantum potential, $Q$, which is required to integrate the hydrodynamic equations (see Eq. 4).  Note that the methodology itself is extremely general and can be used to estimate the probability distribution function (PDF) of any distribution of sample points.

In general, the $N$-dimensional covariance matrices specifying each gaussian cluster reflect the correlation between various degrees of freedom.  A fully specified covariance will require that we solve $N^2$ 
simultaneous coupled equations per gaussian cluster.  
In practice, it is easier to use more gaussians with less covariance than few gaussian clusters with a high degree of covariance.

\begin{figure}[t]
\includegraphics{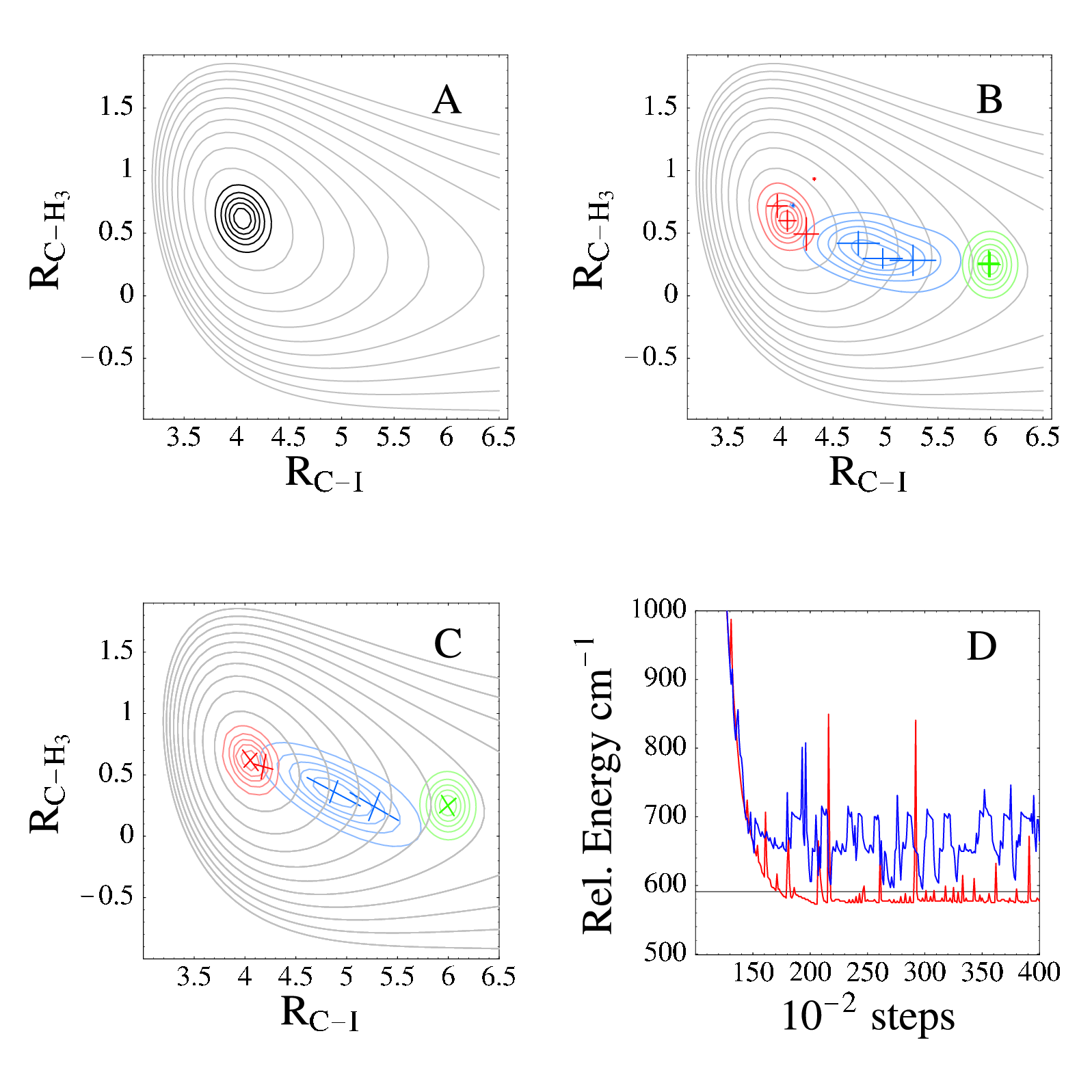}
\caption{Comparison of ground state calculations via exact diagonalization of vs. clustering dynamics for bond stretching modes in methyl iodide.  The vibrational potential energy surface is given as light-gray contours in terms of the 
C-I stretching mode ($R_{C-I}$) and the symmetric ÒumbrellaÓ mode of the methyl hydrogens ($R_{CH_3}$).
 (From M. Shapiro and R. Bershon, J. Chem. Phys. {\bf 73}, 3810 (1980)).  In (a), the red grid shows the location of the DVR points used to converge the ground state and the first few vibrational modes.  A more detailed calculation (and dynamics) requires that this grid cover the entire region of the potential energy surface. In (b), the colored (green, blue, and red) contours show the evolution of the density as it relaxes to the lowest energy state.  Superimposed crosses (+) show the centers and full width at half maximum of the gaussians used to represent the density.  In(c), we show the relaxation to the ground state, this time using two separable Gaussians with full covariance.  In (d), we plot the energy of the system vs. time: \color{blue}{\bf --} \color{black}
case A, \color{red}{\bf --} \color{black}
 case B, \color{black}{\bf --} \color{black}
 DVR calculation. }\label{Fig2}
\end{figure}

Our initial application of this method has been to determine ground states of high-dimensional systems, such as (He)$_n$
 clusters for up to $n=4$. In Fig.~\ref{Fig2}, we demonstrate this by showing the convergence of an initial distribution of sampling points to the vibrational ground state for the anharmonic CH$_3$-I bending/stretching modes of the methyl iodide molecule in its lowest electronic state. This system serves as a useful benchmark for our approach since the vibrational ground state can also be computed by diagonalization of the Hamiltonian matrix. Using this approach gives the ground state energy as 591.045 cm$^{-1}$ above the bottom of the well.  The discrete variable representation (DVR) grid used to converge this and the lowest few vibrational excited states is shown superimposed over potential energy surface in Fig.\ref{Fig2}(a).  In the clustering approach for bound-state problems, the sample points remain localized in a small volume of configuration space as the algorithm relaxes to the minimal energy configuration. In Fig.\ref{Fig2}(d), we show the energy as the system relaxes.  The horizontal solid grey line is the ``exact'' result.  In Fig.\ref{Fig2}(b), our initial sampling is well away from the lowest energy point on the potential energy surface. The superimposed ovals give the location and widths of the $M$ gaussian clusters used in this calculation.  For this calculation we use 4 fully factorized gaussians with no covariance.  As the calculation progresses, the distribution follows the potential energy curve to its minimum and the distribution takes the form of the lowest energy quantum state.  The final relaxed ground-state energy is $665.205 \pm 33.6$ cm$^{-1}$.  At this point, the sample points are still in motion and we are considerably above the DVR ground state energy. Finally, we consider what happens if one takes fewer Gaussians each with full covariance in $(x,y)$.  Starting from the same distribution as before, we propagate the particles applying a viscous force to bleed away the kinetic energy.  Under this approach, the system relaxes to $580.0\pm10.1$ cm$^{-1}$ using only two Gaussians.  Surprisingly, most of the amplitude in the final state is concentrated in the Gaussian located above the minimum in the potential energy surface.  Looking at Fig.\ref{Fig2}(d), we notice that the energy fluctuates about the final average with a relatively small deviation.  However, there are some spikes and these correspond to cases in which one of the clusters, typically the smaller one, suddenly jumped to a different position or the correlation matrix rotated slightly.  Fortunately, the algorithm rapidly corrects for these excursions.    

In many ways, the hydrodynamic formalism implemented with the clustering approach is akin to the guide function Monte Carlo (GFMC) widely used to calculate the ground state of multi-particle systems. For example, in GFMC one chooses an appropriate guide-function from a variational Monte Carlo calculation and uses the guide function to improve the Monte Carlo sampling.  Much like the Bohmian quantum potential, the local kinetic energy determined by the curvature in the distribution function and iterative refinements to the sampling are made using the Monte Carlo algorithm. In our case, both the Òguide functionÓ and the sampling points are determined at each step in the calculation and the sampling points evolve according to hydrodynamic equations of motion.  We expect that this approach will be most useful in high-dimensional problems where conventional variational basis set approaches are numerically impossible.  

\section{  Beyond Bohmian mechanics: Adaptive quantum paths}

\subsection{What kinds of grids and paths are there?}

	Near nodes or nodal surfaces, places where  , the propagation of Bohmian trajectories becomes problematic.  On one hand, as mentioned earlier, these trajectories tend to avoid nodal regions and this leads to an under-sampling problem.  For trajectories near the node, derivative evaluation of the hydrodynamic fields becomes increasingly inaccurate, thus leading to computational breakdown. A closely related problem is that the quantum potential near nodes acquires large values, and small errors in evaluating $Q$ in turn leads to large errors in the positions and momenta for the nearby quantum trajectories.  In order to circumvent these difficulties, we will take a more general look at adaptive dynamic (moving) grids.\cite{ref11}
  The advantage of these grids is the flexibility afforded in directing the grid paths.  Furthermore, these dynamically adaptive grids may be robust enough that accurate results can be obtained over long propagation times. 

When the spatial coordinates associated with hydrodynamic equations of motion are discretized, each grid point follows a time-dependent path, $x_j(t)$, with an associated grid velocity, $\dot{x}_j$. 
The primary advantage of time-dependent grids is that many fewer points may be required because the grid can be chosen to adapt to the evolving hydrodynamic fields. In favorable cases, interesting features may be captured with the moving grid that would require significantly more fixed grid points in order to achieve comparable resolution. There are two sub-categories of moving grids.
\begin{itemize}
\item Lagrangian grids.  For this frequently used type of dynamic grid, the grid point velocity is the same as the local flow speed of the fluid,  .  As a consequence, the grid points march in step with the fluid. Quantum trajectories obtained by integrating the hydrodynamic equations of motion fall within this category (Sec. II. A.)  These trajectories follow along the fluid flow, but an observer has no control over the paths taken by the grid points.
\item Arbitrary Lagrangian Eulerian (ALE) grids.  For these ÔintermediateÕ grids, the grid point velocity is not the same as that of the fluid,\cite{ref11}.  As a result, the grid either advances on the fluid, or the grid may lag the fluid flow.  In this case, it is useful to introduce the slip velocity,  , whose sign may be either positive or negative.  The grid developer has considerable flexibility is creating these Ôdesigner gridsÕ to satisfy certain objectives.  This design issue will be taken up in more detail in the next section. 
\end{itemize}
\subsection{Moving path transforms of the hydrodynamic equations }

For an observer moving along a path $x(t)$, the rate of change of a function  $f$ is denoted   $df/dt$
(this is the ÔtotalÕ time derivative), whereas the time derivative at a fixed point (Eulerian frame) is $\partial f/\partial t$. 
 These two time derivatives are related through the equation:  $df/dt = \partial f/\partial t + \dot{x}\nabla f$, 
where the last term is referred to as the convective contribution.  From this equation, the derivative at a space-fixed point is given by $\partial f/\partial t = df/dt-\dot{x}\nabla f$. 
This equation allows us to transform equations of motion expressed in the Eulerian frame into the more general ALE frames.  	

          When the three hydrodynamic equations given earlier for $\rho$, $v$, and $S$  are transformed\cite{ref8},
 we obtain the new equations (in the first equation, the $C$-amplitude is $C = \ln \rho/2$).
\begin{eqnarray}
\frac{dC}{dt} &=& w \frac{\partial C}{\partial x} - \frac{1}{2}\frac{\partial v}{\partial x} \\
\frac{dS}{dt} &=& w(mv) + L_q(t)\\
m\frac{dv}{dt} &=& mw\frac{\partial v}{\partial x} - \frac{\partial}{\partial x}(V+Q)
\end{eqnarray}
where, for simplicity, a one dimensional problem has been assumed. It is important to note that the slip velocity appears in the first term in each of these equations.  When  $w=0$, the condition appropriate for a Lagrangian frame, these equations revert back to the ÔusualÕ ones that were presented earlier.  However, in their more general form, these equations can be integrated along paths designed to capture features that develop during the course of the dynamics.  How this can be done forms the subject of the next section.

\subsection{Grid adaptation using the monitor function}

	The local hydrodynamic fields surrounding each moving grid point can be monitored with a user designed function $M(x)$.\cite{ref11}
 There is considerable flexibility in crafting the monitor function and in principle different features can be captured
 at different times.  Usually, the monitor captures the gradient and/or the curvature of the hydrodynamic fields, for example, 
$M(x) = 1 + \alpha |\nabla\cdot u|^2 + \beta \nabla^2u$. 
A constant (unity in this case) is added on the right side to prevent the monitor from becoming very small in regions where the fields are relatively flat.  There are many possibilities and the grid designer must decide what features of the fields require monitoring. Once the monitor is introduced, the next issue is how to adjust the grid points so that they can adapt to these fields.  In regions where the monitor takes on large values, we would like for the grid points to come closer together, but in such a way that crossing of paths is avoided. A commonly used way to do this is through use of the equidistribution principle.\cite{ref11} 

	The equidistribution principle (EP) is easily stated for a one-dimensional problem. First, let   $\{M_j\}$ 
denote the values of the monitor at the grid points $\{x_j\}$ , for $j=1,..,N$.  
The end points $x_1$  and $x_N$  are regarded as fixed during the adaption of the internal points. 
Finally, let  $M_{j+1/2}$ denote the average value of the monitor between points $j$ and $j+1$.  The EP then states,
$M_{j-1/2}(x_j-x_{j-1}) =M_{j+1/2}(x_{j+1}-x_j)= {\rm constant}$. 
A large value of the monitor then forces the spacing between adjacent points to be small.  This equation can also be viewed in terms of the equilibration of a spring system, in which the local monitor functions act as the spring constants.  These ÔsmartÕ springs sense features in the hydrodynamic fields.  The spring analogy has been used in the solution of classical fluid problems, and it has also been used recently in the solution of the quantum hydrodynamic equations.~\cite{ref8}
 In order to prevent relatively sudden changes (ÔjerkinessÕ) in the grid, it is useful to consider a refined version of the EP that was first introduced by Dorfi and Drury~\cite{ref12}
				
\subsection{Application of dynamic grid algorithms}
A dynamically adaptive grid was used to study the scattering of an initial gaussian wavepacket from a repulsive Eckart potential (whose shape is similar to that of a gaussian).~\cite{ref7}
When Bohmian trajectories are propagated, the calculation breaks down shortly after ripples start to develop in the reflected wavepacket.  However, much longer propagation time are obtained 
through use of adaption techniques. In this application, the grid points were adapted
using a monitor function designed to capture the local curvature of the wavefunction. The initial wavepacket has a translational energy  $E_{trans} = 4000 cm^{-1}$, the barrier height is  $V_o = 6000 cm^{-1}$, 
and the barrier is centered at  $x_b = 6 a.u.$
This wavepacket was propagated using $N=249$ grid points between  $x_o=-5$ and $x_N=25$.  
The two edge points were held fixed during the entire propagation sequence (Eulerian frame), but more recent calculations have employed Lagrangian edge points.  The internal points were adapted according to the Dorfi and Drury scheme~\cite{ref12}
 that was mentioned earlier.

\begin{figure}
\includegraphics{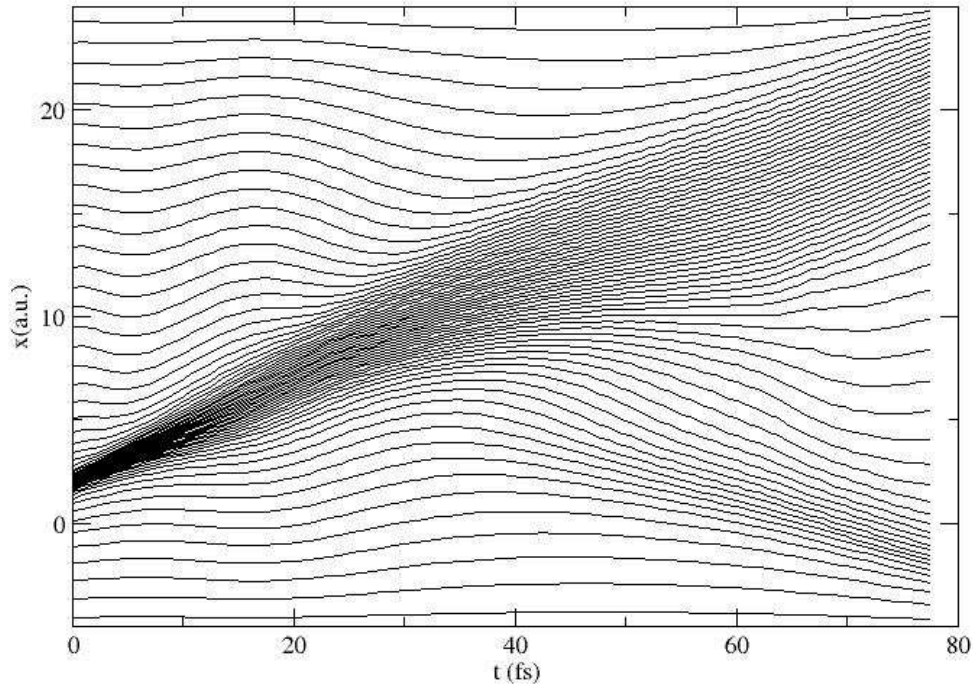}
\caption{Grid paths for wavepacket scattering from an Eckart barrier (only 1/4th of the $N=249$ paths are shown).  
The barrier is centered at  $x_b=6 au)$. Bifurcation of the wavepacket near $t=40fs$ is evident in the clustering of the paths.}
\label{Fig3}
\end{figure}

The paths followed by the grid points (only 1/4-th of which are shown) are plotted in
 Fig.\ref{Fig3}.  At $t=0$, it is seen that the grid points are dominantly clustered in 
the region of large wavefunction density and curvature between $x=1$ and 3. At 
later times, the wavepacket spreads as it moves toward the barrier and this is 
reflected in the grid paths by the width of the clustered region spreading as the 
center heads toward the barrier.  At around $t=40$fs, wavepacket bifurcation is 
manifest by the grid paths diverging near the barrier region to follow the reflected 
and transmitted parts of the wavepacket.  After about 40fs,  these two wavepackets 
are moving toward the lower right and upper right of the figure, respectively. By 
design, the grid points tend to congregate in these two regions of higher density. 

\section{Future studies}

The use of quantum Lagrangian trajectories and more general dynamic adaptive 
grids have the potential to solve quantum dynamical problems in multiple 
dimensions.  The grid points follow the evolving probability fluid so that numerous 
points or basis functions are not wasted in regions of little activity.  These methods 
have already yielded insights into a number of problems in chemical physics, 
including barrier tunneling, electronic nonadiabiatic dynamics, decoherence, and 
relaxation in dissipative environments.

In order to make these methods robust enough to tackle multi-dimensional 
dynamics on anharmonic potential surfaces, additional effort needs to be directed 
toward the following interconnected issues.  (1) We have already mentioned that the quantum potential becomes large around wavefunction nodes and quasinodes and that fluid elements following Bohmian trajectories inflate away from these regions.  
Accurate computation of the quantum potential is very difficult and this in turn can 
lead to instabilities in the equations of motion.  (2)  Maintaining long time stability of 
the solutions is difficult, in part because of the large quantum potentials that arise 
locally.  (3)  Derivative evaluation on the unstructured mesh formed by the evolving 
fluid elements is difficult, although moving least squares or transformation to a 
structured grid for subsequent derivative evaluation are useful strategies.  (4)  Near 
caustics that form when classical trajectories are integrated, the quantum forces 
acting on the fluid elements are both large and rapidly changing.  The hydrodynamic equations are stiff, and special integration techniques (so far untested) are required 
for stable propagation.  All of the unresolved issues mentioned in this paragraph 
need further investigation.

\begin{acknowledgments}
RW and EB both thank the National Science Foundation and the Robert Welch Foundation for financial support.  In addition, many discussions with Jeremy Maddox, Kyungsun Na, Keith Hughes, and Corey Trahan are gratefully acknowledged.
\end{acknowledgments}

\end{document}